\documentclass[sigconf]{acmart}

\usepackage{amsmath,amsfonts}
\usepackage{algorithmic}
\usepackage{graphicx}
\usepackage{textcomp}
\usepackage{xcolor}
\usepackage{multirow}
\usepackage{multicol}
\usepackage{array}
\def\BibTeX{{\rm B\kern-.05em{\sc i\kern-.025em b}\kern-.08em
    T\kern-.1667em\lower.7ex\hbox{E}\kern-.125emX}}
    
\usepackage[utf8]{inputenc}
\usepackage{xcolor}
\usepackage{pdfpages}
\usepackage{graphicx}
\usepackage{enumitem}

\title{``Money makes the world go around'': Identifying Barriers to Better Privacy in Children's Apps From Developers' Perspectives}
\author{Anirudh Ekambaranathan}
\email{anirudh.ekam@cs.ox.ac.uk}
\affiliation{%
  \institution{University of Oxford}
  \city{Oxford}
  \country{UK}
}
\author{Jun Zhao}
\email{jun.zhao@cs.ox.uk}
\affiliation{%
  \institution{University of Oxford}
  \city{Oxford}
  \country{UK}
}
\author{Max Van Kleek}
\email{max.van.kleek@cs.ox.ac.uk}
\affiliation{%
  \institution{University of Oxford}
  \city{Oxford}
  \country{UK}
}

\copyrightyear{2021}
\acmYear{2021}
\setcopyright{acmcopyright}\acmConference[CHI '21]{CHI Conference on Human Factors in Computing Systems}{May 8--13, 2021}{Yokohama, Japan}
\acmBooktitle{CHI Conference on Human Factors in Computing Systems (CHI '21), May 8--13, 2021, Yokohama, Japan}
\acmPrice{15.00}
\acmDOI{10.1145/3411764.3445599}
\acmISBN{978-1-4503-8096-6/21/05}

\begin{document}
\begin{abstract}
The industry for children’s apps is thriving at the cost of children’s privacy: these apps routinely disclose children’s data to multiple data trackers and ad networks. As children spend increasing time online, such exposure accumulates to long-term privacy risks. In this paper, we used a mixed-methods approach to investigate why this is happening and how developers might change their practices. We base our analysis against 5 leading data protection frameworks that set out requirements and recommendations for data collection in children's apps. To understand developers' perspectives and constraints, we conducted 134 surveys and 20 semi-structured interviews with popular Android children’s app developers. Our analysis revealed that developers largely respect children’s best interests; however, they have to make compromises due to limited monetisation options, perceived harmlessness of certain third-party libraries, and lack of availability of design guidelines. We identified concrete approaches and directions for future research to help overcome these barriers.

\end{abstract}


\maketitle

\section{Introduction}
Children are now spending an unprecedented amount of time online~\cite{livingstone2017children}. In the UK, for instance, over 83\% of children between the ages of 12 and 15 own a smartphone, and spend over 20 hours a week using online apps and services~\cite{britain2018children}. It is not only teens and older children who are establishing an increasing presence online; even children under the age of 5 are now spending time daily on tablets and smartphones, leading to an increase in the number of apps designed for children \cite{britain2018children}. 

Children are a vulnerable user group to a wide range of online risks, such as exposure to inappropriate content \cite{chassiakos2016children,gentile2011media,wilson2008media} and potential health risks caused by prolonged screen time \cite{sigman2012time}. However, one particularly important, yet often overlooked, risk for children today is the omnipresence of data trackers in third-party libraries in apps they use every day \cite{binns2018third, razaghpanah2018apps}. These libraries often collect sensitive data about children \cite{book2013longitudinal, lin2013understanding, reyes2018won} including location information, which can then be sent to data brokers for data profiling. Such trackers are widely identified in apps often used by children~\cite{binns2018third}, because developers predominantly rely on targeted advertisements from third-party libraries as their key revenue \cite{leontiadis2012don, acquisti2016economics, interactive2015iab}. Moreover, previous research has shown that only a fraction of the apps categorised to be intended for children are associated with a privacy policy \cite{liccardi2014can}.



In response to concerns about the fraught state of children's privacy, governments and public sector organisations have assembled working groups and consultations to understand the privacy landscape for children in the digital space and proposed regulatory interventions \footnote{\textit{See Call for evidence – Age-appropriate design code}
(https://ico.org.uk/about-the-ico/ico-and-stakeholderconsultations/call-for-evidence-age-appropriate-designcode/).}. The privacy landscape has seen significantly changed in the past few years. For example, in 2018 Europe saw the introduction of the General Data Protection Regulation (GDPR), which aims to recognise personal data as a fundamental right, which was followed by a specific section about GDPR for children (GDPR-K). Another important initiative is the statutory Age Appropriate Design Code issued by the Information Commissioner's Office (ICO) in the UK \cite{ico_2020}, as a clarification of GDPR-K in the UK, aiming to make data protection a key element when designing services for children from 2021. 

Despite these initiatives, research has shown that children's data protection still has a long way to go. For example, recent research shows that developers still rely on privacy invasive advertising networks in their development practices \cite{mhaidli2019we}.%
Thus in this work we aim to examine the following research questions:
\begin{itemize}
    \item RQ1: What did app developers perceive as their responsibilities when designing for children?
    
    \item RQ2: What are the main practices adopted by developers for achieving data protection and how do they align with the leading data protection frameworks from different sectors?
    
    \item RQ3: What are key barriers for the adoption of the leading data protections for the app developers and how can these barriers be overcome?
\end{itemize}

We conducted 20 interviews and 134 surveys with family app developers, and accompanied by an analysis of 5 leading data protection frameworks to understand best practices expected from developers and how they support developers in implementing these practices. Our findings show that developers feel responsible for designing apps with the best interests of children in mind, including that children's privacy should be respected and thus data collection should be minimised. However, we also identified several barriers which were in the way of realising best practices set out in the data protection frameworks. 
First, developers find it difficult to navigate the complex and opaque landscape of third-party libraries. There is a lack of awareness of age-appropriate libraries, forcing developers to choose major data controllers as their third-party service providers. 
Second, developers struggled to monetise their apps in ways that did not rely on advertising. The market pressure and competition made alternative and potentially privacy-friendly business models, such as offering premium apps, unsustainable. 
Third, there was a general lack of awareness of guidelines for designing for children, resulting in developers relying on terms and requirements set out by market leaders, such as Google, which are not always aligned with the best interests of children. 

We found that data protection frameworks did not sufficiently support developers in addressing these barriers. Based on our findings, we make recommendations to help improve the state of app development for children. We propose to include concrete recommendations in guidelines regarding selecting data controllers, advertising networks, and third-party libraries. Furthermore, we also advocate for heightened support for both end users and developers through tools which can shed light on the otherwise opaque data economy. Lastly, our research also highlights the importance of needing industry support to make major changes to the app development landscape.

\section{Background}

\subsection{The Importance of Children's Privacy}
Children spend more time on digital devices now than ever before \cite{ofcom2018children} and this is raising concerns in public and policy circles about children's privacy and the commercial use of their data \cite{lupton2017datafied}. As it stands, children's privacy finds itself in a vulnerable and endangered position \cite{livingstone2017children}. In the UK alone, 82\% of children between 5 and 7 spend almost 10 hours per week online \cite{ofcom2018children}. This is particularly significant today, as the digital platforms children use and its resulting data traces are owned by the private sector \cite{montgomery2015youth}. In fact, commercial organisations are gathering more data from children than governments are capable of \cite{nyst_gorostiaga_geary_2018}. They use a range of methods, often invasive, to track children's activities \cite{montgomery2017children}, such as cookie-placements, web-beacons, and advertising IDs. In addition, children are also often nudged into disclosing more personal information than is necessary \cite{bailey2015perfect, shin2016adolescents} or as a trade-off to access a service \cite{micheti2010fixing, lapenta2015youth}. 

However, despite their omnipresence online, children do not understand digital risks as well as most adults \cite{livingstone2017children}, and they have a poor understanding of privacy related contexts \cite{zhao2019make, kumar2017no}. For example, children fail to understand why their data is valuable to third parties \cite{lapenta2015youth}, how their data is collected \cite{emanuel2014exploring, acker2018youth}, how it is stored and analysed  \cite{bowler2017lives}, and how it may be used in the future \cite{murumaa2015drawing, bowler2017lives, pangrazio2018s}. They also find it challenging to understand privacy terms and conditions, because of their length and legalese \cite{best2017growing}, and feel forced to accept the terms laid out to them \cite{lapenta2015youth}. Children feel that targeted advertising and profiling is a part of digital life \cite{lapenta2015youth} and that there is very little they can change about their behaviour to prevent this \cite{lapenta2015youth, pangrazio2018s}. 

Loss of privacy and data sharing to third parties is known to lead to concrete harms, such as identity theft and fraud \cite{coughlan2018sharenting}, and the normalisation of a data surveillance culture \cite{kobie2016surveillance}. However, the most critical reason why a proactive stance against data collection is needed, is due to harms arising from long term risks to children's reputation and opportunities as they grow older \cite{longfield2018knows}. These harms and risks are particularly important to consider, as their exact nature is still unknown and may evolve over time.

It is for these reasons that children form a particularly vulnerable user group and that concerns have been expressed about the the `datafication' of children \cite{lupton2017datafied}. The need for children's support in navigating privacy choices, is not entirely ungrounded. In the next section we discuss the pervasiveness of data harvesting features in apps for children. 





    


\subsection{Why Children's Privacy is at Risk}
Mobile apps have been shown to be particularly threatening to children's privacy \cite{binns2018third}. This is primarily attributed to the use of third-party libraries, which are increasingly prevalent in today's apps \cite{balebako2014privacy}.

These libraries have permissions to collect sensitive data \cite{book2013longitudinal, lin2013understanding} and frequently access location permissions. They are known to track call logs, browser history, and contact information for the purpose of targeted advertisements, even if that was not the intended functionality \cite{grace2012unsafe}. This is not different for children's data, as apps in the ``Family" category of the Google Play Store have shown to have the second highest number of data trackers associated with them \cite{binns2018third}. The prevalence of third-party libraries can be explained by the fact that developers rely on targeted advertising for generating revenue \cite{leontiadis2012don, acquisti2016economics, interactive2015iab}, which in turn uses third-party libraries to collect targeted data. Additionally, they also simplify development, provide increased functionality, and may be more secure than proprietary software modules \cite{enisa2018}.

Addressing these trackers is not trivial on mobile applications, as Android and major other smartphone operating systems do not provide end users the freedom to control third-party tracking through apps. Users are therefore dependent on privacy regulations and app review processes of marketplaces \cite{anderson2010inglorious} to protect them. However, review processes are not always transparent, and Google and Apple have thus far been poorly incentivised to exert control over the data tracking ecosystem, as they hold a vested interest in the advertising industry \cite{google_advertising}. 

Instead, we have seen an increase in tools to better inform users on the dissemination of their data to third-party trackers \cite{balebako2013little,chitkara2017does,srivastava2017privacyproxy,van2017better}. However, more often than not, such tools are aimed at adults, and both children and adults may have a difficult time contextualising implications of tracking in the privacy and security landscape.

\subsection{Regulatory Interventions}
The problem in tackling these issues that children's privacy rights have thus far not been sufficiently supported \cite{livingstone2016one} and there have been calls for regulatory interventions to address this issue \cite{kidron2018disrupted}. In recent years, the development landscape has seen changes. Europe introduced the General Data Protection Regulation (GDPR) in 2018, which recognises the protection of personal data as a fundamental right. The UK Information Commissioner's Office put into effect a statutory code for developers requiring them to make data protection a central tenet in their design \cite{ico_2020}. It is not only public bodies which have pushed for changes, in 2019 Apple changed their policies to prohibit third-party advertising and analytics \cite{apple_iphone_facebook}.

\subsection{Privacy and Security Practices of Developers}
The privacy changes seen in recent years is placing more responsibility on developers to create appropriate apps for children. However, research on  developers' privacy perceptions and practices is limited.  In a survey with 228 app developers exploring security and privacy decisions, it was shown that developers often are not familiar with the practices of third-party APIs due to difficulties in reading their privacy policies \cite{balebako2014privacy}. Another study investigated the app developers' preference for using advertising as a revenue model and how they selected advertising networks \cite{mhaidli2019we}. The study found that developers perceived advertising often as the only way to profit off apps and often chose advertising networks based on their popularity.

\begin{table*}[t]
\caption{\label{table_ico_codes} A summary of the data protection codes in the AADC.}
\small
\centering
\begin{tabular}{p{1.2in}p{4in}}
\hline
\textbf{AADC code} & \textbf{Definition}  \\ \hline
Code 5: Detrimental use of data & Does not exploit children's data for purposes that may harm their health or wellbeing. \\ \hline

Code 8: Data minimisation & Collect and retain only the minimum amount of personal data you need to provide the elements  of your service in which a child is actively and knowingly engaged. Give children separate choices over which elements they wish to activate.  \\ \hline

Code 9: Data sharing & Do not disclose children’s data unless you can demonstrate a compelling reason to do so, taking account of the best interests of the child. \\ \hline

Code 10: Geolocation & Switch geolocation options off by default (unless you can demonstrate a compelling reason for geolocation to be switched on by default, taking account of the best interests of the child), and provide an obvious sign for children when location tracking is active. Options which make a child’s location visible to others should default back to ‘off’ at the end of each session. \\ \hline

Code 12: Profiling & Switch profiling ‘off’ by default (unless you can demonstrate a compelling reason for profiling to be on by default, taking account of the best interests of the child). Only allow profiling if you have appropriate measures in place to protect the child from any harmful effects (in particular, being fed content that is detrimental to their health or wellbeing). \\ \hline
\end{tabular}
\normalsize
\end{table*}

\section{Children's Data protection frameworks}
In this section we present our review of five leading children's data protection frameworks from three different sectors (regulatory, private, and human rights organisations), with the aim of understanding requirements and expectations put on developers. Given that the main objective of our research is to investigate app developers' data protection practices, our review focuses on parts of the frameworks related to data collecting and handling.

To compare across these frameworks, we use the statutory Age Appropriate Design Code (AADC), developed by the Information Commissioner's Office (ICO) in the UK as the benchmark framework, which is seen as one of the most comprehensive regulatory frameworks to date and which fits with our analysis apps from the UK app market. We aligned the data protection principles from the AADC against those from the following four frameworks: 
\begin{enumerate}[noitemsep]
    \item \textit{(Regulatory)} COPPA - The US Children's Online Privacy Protection Act;
    \item \textit{(Human rights)} COPFE - UNICEF's Children’s Online Privacy \& Freedom of Expression ;
    \item \textit{(Private sector)} Google's `Designing Apps for Children and Families';
    \item \textit{(Private sector)} Apple's App Store Review Guidelines.
\end{enumerate}

We have not included the GDPR(-K) in this, because the AADC is designed to be a clarification of the enforcement of GDPR-K.



\subsection{The UK ICO Age Appropriate Design (AADC)}
The UK Information Commissioner's Office (ICO) is an independent authority aimed at protecting and upholding information rights in the public interest and promoting data privacy for individuals. In an effort to address privacy concerns of children in the digital world, the ICO introduced the statutory Age Appropriate Design Code (AADC) \cite{ico_2020} \footnote{The latest ICO Code for Age Appropriate Design is available at \url{https://ico.org.uk/for-organisations/guide-to-data-protection/key-data-protection-themes/age-appropriate-design-a-code-of-practice-for-online-services/5-detrimental-use-of-data/}; retrieved on 4 August 2020.}. The code aims to ensure online services safeguard children's personal data and comply with the GDPR. If services are not compliant by September 2021, the ICO can issue firms enforcement notices and fines up to 4\% of their global turnover if they breach these data protection guidelines. 


The AADC consists of 15 codes which take into account principles set out in the United Nations Convention on the Rights of the Child (UNCRC). The code touches on many different aspects related to design for children, including user-facing design practices (such as `transparency', or `high privacy by default'), alignment with fundamental principles (such as supporting children's best interest), support for data protection in new emerging technologies (e.g. connected toys), and regulations regarding data collection and minimisation. 

The focus of our study is to understand children's app developers' choices around data handling practices, and therefore we focus on those AADC codes specially related to data collection and minimisation. We excluded codes not directly related to development practices or data handling, such as `Data protection impact assessments' and `Transparency'. We identified five codes from the AADC that are directly related to regulations of processing children's data, which are summarised and explained in Table~\ref{table_ico_codes}. 







\begin{table*}[t]
\caption{\label{table_frameworks} This table lists 4 leading data protection frameworks and their alignment with the data protection principles set out by the AADC. The cells without content indicate that the framework does not mention the associated AADC code. }
\small
\centering
\begin{tabular}{|p{0.12\textwidth}||p{0.2\textwidth}|p{0.2\textwidth}|p{0.2\textwidth}|p{0.15\textwidth}|}
\hline
         \textbf{AADC Codes}  & \textbf{UNICEF} & \textbf{COPPA}  & \textbf{Apple App Store} & \textbf{Google Play}  \\ \hline
\emph{5: Detrimental use of data} & Children have the right not to be subjected to attacks on their reputation &  - & -  &   -    \\ \hline

\emph{8: Data minimisation} & Children’s data are kept to what is minimally necessary & Should not condition a child's participation on the collection of more than the personal information that is required to drive that activity  & Only access to data relevant to the core functionality of the app is permitted  & The collection of any personal and sensitive information must be disclosed    \\ \hline

\textit{9: Data sharing}  &  Refrain from sharing information that could undermine children’s current or future reputation & Disclosure to third parties requires parental consent and assurance that reasonable security measures are in place & Prohibited from sending PII or device information to third parties; this includes IDFA, location, and device information   & -     \\ \hline

\textit{10: Geolocation }       &   -     & Geolocation is treated as PII and needs parental consent before collection;  & If directly relevant to the features and services provided by the app  & Cannot request location permissions       \\ \hline

\textit{12: Profiling }      & Children enjoy protection from online profiling         &  The use of any persistent identifiers for the identification of a specific individual is prohibited           &  Third-party analytics or third-party advertising prohibited   & Personalised advertising prohibited       \\ \hline

\end{tabular}
\normalsize
\end{table*}


\subsection{UNCRC}


The United Nations Children's Fund (UNICEF), is an agency which is part of the United Nations and globally provides humanitarian aid to children. To support developers in realising the fundamental privacy rights, in 2018 UNICEF introduced the `Children’s Online Privacy \& Freedom of Expression' (COPFE) industry toolkit, which describes five overarching principles~\cite{unicef_tools, nyst_gorostiaga_geary_2018} to protect children's right to privacy, personal data freedom of expression, protection of reputation, and access to remedy. 

This framework is largely aligned with the AADC, with a great emphasis on children's fundamental rights; however, the framework also discusses many additional elements on the rights for children, such as the rights to be educated, have access to resources for risk coping, and the right for parents to have access to resources to help their children. Five of the sub-codes are identified to be closely related to the set of AADC benchmark codes, with some slight different emphases.

\textbf{Data profiling} is discussed in COPFE, at a more general level: ``Children enjoy protection from online profiling''. It does not specifically give examples of profiling or explain what constituted profiling. In the checklist, profiling is placed in the context of behavioural advertising and suggests applying \textit{``specific protection''}, as it \textit{``involves collecting and aggregating personal data''}. As for \textbf{data minimisation}, COPFE states that collected data should be \textit{``fit for purpose''} and limited to \textit{``what is minimally necessary''}. The stakeholders involved in \textbf{data sharing} are explicitly named to include both \textit{``parents or guardians''}, \textit{``media outlets and other third parties''}, who should \textit{``refrain from sharing information that could undermine children’s current or future reputation''}. Lastly, COPFE also argues against \textbf{detrimental use of data}, stating that ``children have the right not to be subjected to attacks on their reputation'' and ``can seek the removal of content they believe is damaging to their reputation''. It does not explicitly talk about geolocation or location settings.

\subsection{COPPA}
The US Children’s Online Privacy Protection Act COPPA Rule was established primarily to curb direct marketing aimed at children under 13. It first went into effect in 2000 to implement the US Children’s Online Privacy Protection Act (COPPA) 1998. COPPA is aimed at operators who provide services targeted to children under the age of 13. 

Broadly speaking, COPPA focuses on the enforcement of a clear online privacy policy for services targeted at children, and supporting parents to provide consent on their children's behalf and protect their children's personal data online. COPPA has one specific rule about data handling, which describes that children's data should be retained \textit{``for only as long as is necessary to fulfill the purpose for which it was collected and delete the information using reasonable measures to protect against its unauthorized access or use''} \cite{ftc_faq}.  

COPPA has an inherent \textbf{data minimisation} principle: service providers \textit{``Should not condition a child's participation on the collection of more than the personal information that is required to drive that activity''}. It has no specific conditions on \textbf{data sharing}, apart from that it requires parental consent assurance that reasonable security measures are in place by the party the data is being shared with. \textbf{Geolocation} data is treated as personally identifying information if it can be narrowed down to street or city name, in which case parental consent is requried. Regarding data \textbf{profiling}, a service is COPPA-compliant so long as no \textit{``specific individual can be identified profiling including through their persistent identifiers''}. Under COPPA, the use of any persistent identifiers for the identification of a specific individual is prohibited.

\subsection{Private Sector Guidelines - Apple and Google}
When publishing apps on Google's Playstore or Apple's Appstore, there are specific terms that the app has to uphold. Both Google and Apple have sections reserved for apps targeted at children in these terms. 

Google has these outlined in their `Designing Apps for Children and Families' \cite{google_guide} guidelines. Apps targeted at children must comply with the `Designed for Families' programme, which lists 12 requirements apps must fulfil. The requirements cover different aspects of app design, including the content of apps, policies regarding ads, and special restrictions about the use of Augmented Reality. Advertising is only allowed through certified ad SDKs \cite{google_ads} and personalised advertising, where users' behaviour and interest data is used to customise advertising content, is not allowed. Collecting personal data of users is allowed as long as it is disclosed to them. Google further requires apps to comply with GDPR, COPPA, and other applicable regulations.

Similarly, before publishing on Apple's App Store, apps have to comply with the `App Store Review Guidelines' \cite{apple_guide}. Apple has specific sections in their terms aimed at protecting children's privacy. Third-party analytics and third-party advertising for children is permitted in limited cases, given that services do not transmit personally identifiable information and have their practices and policies publicly documented. 


Neither guidelines state anything specifically regarding \textbf{detrimental use of data} in the context of data processing, while both have clear requirements about \textbf{data minimisation}. Apple requires apps to only request access to data relevant to the core functionality of the app \cite{apple_guide}, and Google requires apps to disclose the collection of any personal and sensitive information \cite{google_guide}. This is also the same for \textbf{profiling}, where neither allow personalised advertising and marketing. Apple also does not allow third-party targeted analytics and third-party advertising. In fact, Apple does not allow any \textbf{data sharing} with third parties, while Google does not have a clear policy on this. 
\textbf{Geolocation} is also discussed in both guidelines. In this regard, Google is stricter than both the AADC and Apple, requiring that \textit{``Apps designed specifically for children cannot request location permissions''} \cite{google_guide}. Apple's stance is more aligned with the AADC and requires that location services should only be used \textit{``when it is directly relevant''} \cite{apple_guide}.  



\subsection{Analysis Summary}

Table \ref{table_frameworks} summarises how the terms specific to data handling in these guidelines are aligned with the AADC codes previously presented in Table~\ref{table_ico_codes}. For those principles which do agree with AADC, we included the definition of the principle to highlight the nuanced differences between all the frameworks. 

It shows that regulation/guidance around data-based \textbf{profiling} is required by all frameworks; however, there are some nuanced emphases amongst them. The AADC takes the position that children merit specific protections with regard to the use of their personal data, and therefore requires profiling to be transparent and to be turned off by default. COPPA is stricter in this regard, prohibiting any profiling that may identify any specific individual. Apple and Google are aligned with this, and prohibit personalised advertising and marketing, with Apple stricter than Google and the AADC, as they prohibit third-party analytics and advertising. 


\textbf{Data minimisation} is also widely required by all frameworks, but with different emphases. Apple's is most aligned with the AADC, although it does not provide an option for children to `make separate choices over which elements' to be activated yet. Interestingly, because of its sensitive nature, \textbf{geolocation} is specially discussed in all the frameworks except COPFE. While Apple allows the use of geolocation with reason, Google does not allow location permissions to be requested in any app directed at children. 


Finally, \textbf{detrimental use of data} is only discussed by COPFE and AADC, as probably both have a primary focus on children's wellbeing and best interest. It can be interesting to see with the effect of AADC in September 2020, how this may effect the current guidelines.


The analysis of the five leading children's data protection frameworks gives us a concrete understanding of the current landscape for protecting the collection, processing and handling of children's data.  We will revisit this framework with the findings from our studies with children's app developers for Google's Playstore, and identify the gaps and barriers for the implementations of these guidelines.

\section{Methods}
In this work, we seek to understand how developers perceive their responsibilities in the creation of apps for children, and how such perceptions impact app development practices. To this end, we conducted IRB-approved semi-structured interviews and surveys with them.

The interview and survey included similar open-ended questions to allow developers to process the ideas before elaborating on them during the interview. They were designed to capture developer perceptions of risks online for children, their views on data collection practices by third-party libraries and trackers, and their development practices. Below are sample questions which the survey and interview had in common: 

\begin{itemize}[noitemsep]
    \item What do you think privacy and security risks are for children online and which role do parents play in this? 
    \item What are your views on current data collection practices used by apps and third-party companies right now?
    \item What are your practices when it comes to using third-party libraries and how do you ensure its safety?
    \item What development process do you follow while developing the app? Do you follow any known design or development process?
\end{itemize}

The full survey is available in the supplementary materials.




\subsection{Surveys}

\subsubsection{Participants}
Participants were recruited through direct email communications using the address they made publicly available on the Google Playstore. We contacted developers whose apps were declared from the `family’ and `education’ genres (not excluding apps with particular age ratings), available in Europe. We then extended this to developers of parental control apps, as they are increasingly used by families and children. They were retrieved using keyword search ~\cite{wisniewski2017parental}, including terms such as `parental control', `online safety' and `online privacy'. We sent out a total of 11,000 emails and received 134 survey responses (S1-S134).

\subsubsection{Procedure}
The online survey was distributed in late July 2019 to participants who had agreed to join in the study. The survey was designed to take approximately 15 minutes. Before beginning the survey, participants had to give consent, indicate that they were over the age of 18, and that they had read the accompanying information sheet. The survey included closed, open-ended, short answer, and likert-scale questions. Participants were not compensated for completing the survey. The study was approved by the research ethics committee of the university.

\subsection{Interviews}
\subsubsection{Participants}
We recruited interview participants using the same method as described above and conducted interviews with 20 family app developers. The interviews were held remotely between July and November of 2019. Participants were not compensated for participating in the interview. 



\subsubsection{Procedures}
All interviews were conducted remotely over Skype and participants consented to being audio recorded. Interviews lasted between 24 and 43 minutes. 

Participants were initially asked to provide background information about their development experiences, the motivation to develop their app, and its functionality. Then, we asked questions about 1) their perceptions of risk for children online; 2) their views on data collection practices in the mobile ecosystem; 3) their personal data collection practices in their app development; and finally 3) current practices of making use of third-party libraries, including any information factors, quality assessment or decision making procedure involved in choosing and enclosing these libraries in their app. 


\subsection{Analysis of Qualitative Data}
The audio recordings were transcribed with all personally identifiable information anonymised. The transcripts and the open-ended survey questions were analysed in two iterations. First, using (coding reliability) Thematic Analysis~\cite{braun2018}, the first two authors coded six different transcripts independently and then convened to consolidate themes to derive a common codebook. Then, using this codebook, the next 8 transcripts were independently coded until an inter-rater reliability score, using Cohen's kappa \cite{cohen1960coefficient}, of 0.85 was achieved. The first author then re-coded the first six transcripts, coded the remaining six transcripts, and coded the open-ended survey questions. 



\subsection{Descriptive Statistics}
\subsubsection{Surveys}
88\% of the survey participants identified as male and 27\% of the participants was between 30 and 39 years old, representing the largest age demographic. 25\% of the participants was between 40 and 49, 22\% was between 23 and 29, 15\% percent was between 18 and 22, and 11\% of the participants was over the age of 50. English was the native language to 26\% of the participants. Most of survey respondents originated from Europe (40.5\%) and Asia (34.4\%), with others from North America (10.7\%), South America (9.3\%) and Africa (4\%).

\subsubsection{Interviews}
The total duration of the interview recordings is 698 minutes. The average interview length is 34.5 minutes (Sd = 6.5). The shortest interview was 24 minutes and the longest 43 minutes. The app characteristics and developer demographics are summarised in Appendix 1.

\section{Results}
In this section, we report our qualitative and quantitative results that provide us an insight about children's app developers' data protection practices, situated in the context of their app development motivations and awareness of current guidelines and regulations. 

Our thematic analysis of the data shows that there have mainly been four major themes: (i) developers' perceived responsibilities and motivations in designing for children's best interests; (ii) perceptions of data collection practices; (iii) reliance on third-party libraries; and (iv) the need to earn money. A summary of results can be found in Table \ref{tab:themes}.

We make use of the descriptors `S' and `I' to refer to the surveys and interviews respectively. 




\subsection{Design for Children's Best Interests}
The majority of developers reported that designing apps for children comes with moral responsibilities (15/20) and they believed that apps for children should be designed `differently' than apps for adults. More specifically, developers emphasised the responsibility to make apps age-appropriate and take children's developmental needs into consideration (10/20):

\begin{quote}
    \textit{``So we interact with the child, as the app interacts with the kid during a very sensitive period, when all the behavioural patterns are formed for the future. So compared to general app, or an app that an adult can use, we think we bear a much more responsibility for the user.''} -- I10
\end{quote}





However, when developers sought to design age-appropriate apps for children, they faced a range of challenges. More than half indicated that they struggled to find suitable and specific design guidelines (11/20). Guidelines provided by the app marketplaces provide limited discussions on how to design for children's developmental needs:

\begin{quote}
    \textit{``There are some general guidelines, like the posts which say, above a certain age you are allowed to show violence, but nothing specific to the education process itself: not pedagogical recommendation, no tutorial design recommendations, nothing of the sort. We wouldn’t mind if such things existed and were paid attention to.''} --- I11
\end{quote}

Developers also indicated that following guidelines has not always been possible, as they constantly changed, and provided insufficient support for smaller companies or individual developers (7/20).


\begin{quote}
 \textit{``We are aware of codes, but they are constantly - the guidelines - they are constantly changing, and our problem is, since we are in another country, here the laws are different. We usually get and try to read every guideline and everything since we are only two people and we don’t have any lawyers or any law advice from lawyers, we sometimes can miss few guidelines.''} -- I9
\end{quote}


As a result, they defaulted to following Google's Guidelines on ``Designing Apps for Children and Families'' (11/20), as it was the most ``accessible'' or ``required'' (10/20). However, the principles from these guidelines referred to are primarily centred around app publication requirements, such as the use of specific ad SDKs: 

\begin{quote}
    \textit{``Google Play wants to be sure about if you wanna advertise a parental control or child oriented application, it is much easier to obey these rules. You have to accept these restrictions to publish your applications''} -- I1. 
\end{quote}



In the end, only a small number of developers used specific methods or techniques for designing for children. A few survey respondents (3\%) and interviewees (3/20) consulted professionals to help them with the app design:


\begin{quote}
    \textit{``We have a very good network of scientists who take a look at what we are doing and steer us in the right direction. For example, we are working together with a professor who knows about a lot of gamification.''} -- I7
\end{quote}

\subsection{Perception of Data Collection Practices}
Survey respondents generally believe that the privacy of children should be respected, in terms of data collection, data sharing, and third-party analytics. The majority of them reported to \textit{not collect personal data} of the users (90\%) and almost all developers indicated to \textit{not share data with third parties} (95\%). Similarly, they strongly felt that it is unethical, even for the purpose of sustaining the business, to sell children's data (90\%) or make use of targeted advertisements containing third-party trackers (85\%):

\begin{quote}
    \textit{``I was approached by about 10 companies to add tracking libraries to my app in exchange for a payment of about 100\$ to 400\$ per month from each company. Since I find this unethical I refuse to do so by so far.	''} -- S11
\end{quote}

Some interviewees were opposed to any data collection in apps due to the associated risks  and potential privacy problems (8/20).

\begin{quote}
    \textit{``I strongly oppose to data collection, especially if its an app directed at children, I don’t think any data should be collected. I think it should be turned off. And there should be very strict policies on that.''} -- I17
\end{quote}

However, a few interviewees stated that limiting collection to non-personal data is not necessarily harmful (4/20) and a small number of them even said that data collection does not affect individuals at all (2/20).

\begin{quote}
    \textit{``Its not always bad. For example, we have to collect some data for our own analytical purposes to design the app better, and these are oftentimes anonymised data; these are not personal data to children.''} -- I12
\end{quote}

\begin{quote}
    \textit{``On an individual basis, I can see this (data collection) being very damaging. [...] But on a mass scale, I just don’t see the risks and the likelihood of the risk being that great.''} -- I5
\end{quote}


Apart from their beliefs about data collection, they were also faced with technical constraints (7/20), such as increased costs of hosting a storage server, and liability issues (4/20) in collecting data:

\begin{quote}
    \textit{``Initially, [we] collected a lot of additional data, like IP addresses of users when they uploaded things, published geographic location, and so on. [..] With the European, you know, all the law changes and so on, we decided okay, lets not do that. We’ll only collect the absolute minimum that we can.''} -- I6
\end{quote}

\subsection{The Need for Analytics and Other Third-Party Libraries}



Developers also mentioned several important goals which could only be realised through third-party services (13/20). For example, some of them expressed the need to further understand user behaviour and interests to improve the app performance and usability (9/20):


\begin{quote}
    \textit{``So when a child [uses our app], it gets a dedicated individual training. And in order to deliver something like this we need to collect training data from the child. But only in order to improve the training for the children.''} -- I7
\end{quote}

Two developers also indicated that children's data is not very ``reliable'' (I1), as children's behaviour is erratic, because ``they don’t even know what they are doing on the phone'' (I1). Another developer explained that children's data does not serve a useful function: 

\begin{quote}
    \textit{``I don’t see the value in it [collecting data] as much as adults, because adults have a lot more capital to spend. So... to be honest with you. So I see the privacy issues on one side, on the other side I don’t see why you’d even want to capture information of people who don’t even have a lot of money.''} -- I5
\end{quote}
At the same time, developers realise that data-based service optimisation is a sensitive topic, especially where it concerns children and happens through third-party services. 

Others stated they needed third-party services to simply maintain efficient core services for their apps (5/20), e.g. \textit{``We use like Firebase for push notifications''} (I4). This was also reflected in the survey data, as the majority of the developers reported to make use of third-party libraries (63\%). The primary reasons for this were `Convenience' (72\%), `To reuse existing software modules' (52\%), for `Targeted ads' (21\%), and to ``Gain insight into user behaviour'' (S53) (6\%).



In the trade-off between collecting data to provide core services and avoiding data collection, developers used various strategies to support children's privacy needs and minimise data collection. Survey respondents indicated to avoid collecting unnecessary data (68\%), put privacy settings high by default (37\%), and did not nudge users into making privacy reducing choices (31\%). Similarly, developers reported to avoid collecting personal data (6/20) and limit themselves to collecting anonymous data without affecting `individuals' (9/20): 


\begin{quote}
    \textit{``I don't collect any children information; we use Firebase Analytics for app usage statistics, which is required for product improvement.''} -- I2
\end{quote}

\subsection{Selecting Third-Parties}


Beyond the conflicting views developers had about the need and appropriateness of data collection for children, they also faced challenges in selecting data controllers and third-party service providers. In particular, the opaqueness of the data economy made it extremely challenging to navigate the APIs and SDKs that are available on the market (13/20). For example, it was unclear to developers what libraries are doing in the background (10/13) and what organisation are secretly doing with the collected data (3/13):


\begin{quote}
    \textit{``We were in this situation where we were so convinced that we don’t show [ads] and don’t do anything wrong. I mean, we didn't \emph{knowingly} gather any data, we didn't run any campaigns, we didn't use any marketing techniques or anything. So there was nothing of that [data colletion] in our ads. [...] For example, Unity, which we use to make our apps,  they have this solution to deliver in-app purchases. So we use it. Now we found out that whenever there is a purchase, Unity takes some personal data, to make this transaction possible. So, and its not cool with us, but there is no other way to do this.''} -- I8
\end{quote}



Developers expressed that they wished data controllers would be more transparent and accountable, but doubted that current governmental regimes would be likely to help achieve this (6/20):


\begin{quote}
    \textit{``Well, I’d love if companies took a little bit more responsibility. But they are not going to; I mean they don’t think twice of increasing user rate. [..] It doesn't seem that politicians nowadays have a vague understanding of how the internet works, or how digital works. And it seems for me to be impossible for them to impose the correct laws.''} -- I17
\end{quote}




Survey respondents reported to have no internal policies regarding the choice of libraries or third-parties. The majority of them chose libraries themselves (39.3\%) or discussed this amongst colleagues (26\%). A small number of participants needed approval from their manager or supervisor (16\%).

In making this choice, developers were left with relatively little information to base their decision.  Thus, they resorted to relying their perceptions of the trustworthiness and quality of the software typically produced by third-parties as a basis for selecting them (11/20):  

\begin{quote}
    \textit{``I trust Google. I used a lot of their tools, and I think they are well developed and are maintained constantly. I use Facebook SDKs and its also very good and well developed library.''} -- I14
\end{quote}

Such providers were also the most popular and prominent, which further made it likely they would be selected by developers. Thus, unsurprisingly, the majority of both interviewees (13/20) and survey respondents (64\%) selected libraries from Google. Survey respondents further indicated that \textit{``Google libraries should be safer than any others''} (S29) and that they \textit{``only select reputable libraries from large companies''} (S113). As a result, 65\% of the survey respondents and 9/20 interviewees reported to use Google's libraries by default, such as Google Analytics (59\%S) and Firebase (22\%S).


\begin{quote}
    \textit{``We are typically using libraries from Apple of Google. Rarely do we use like something we found on a developer network forum, where you could potentially get in trouble. So yeah, the libraries we use are primarily from major parties that have created their own stuff.''} -- I3
\end{quote}

Only a few survey respondents indicated to look at the security aspects, for example by performing `Pentesting' (4\%), `Internal Testing' (1\%), reading the Terms of Service and Privacy Policies (6\%), or considering any reported security issues (4\%): \textit{``We read the terms of use and if it a code library, we do a quick code review to identify any security vulnerabilities''} (S47). 








\subsection{The Need to Make Money}

Understandably, sustaining a viable business was one of the primary objectives for developers to support sustained development (13/20), because by the \textit{``end of the day you've gotta make money. Money makes the world go around''} (I3). This sentiment was reflected by the majority of survey respondents as well, who reported that commercial success of the app was an important factor in development (60\%). In fact, 63\% said they relied on app development as a major source of income, either through the revenue the software generated (37\%), third-party investments (16\%), or through contract work (10\%). 
As a result, the majority (79\%) of them monetised their app, either through ads (32.6\%), a pay-to-download model (25\%), in-app purchases (21.2\%), or subscriptions (18.9\%). Approximately a fifth of the developers, however, did not make use of any monetisation features (21\%) and were running the app for `non-profit', stating to \textit{``develop for fun and learning''} (S46).

Almost all interviewees who did monetise their apps wanted to do this in an age-appropriate way. They felt a moral obligation to not make use of harmful techniques (17/20), such as the use of \textit{``manipulative flashing lights, button that move, sparkling colours''}, because then \textit{``of course [a] child would click on in app purchases: they would get pleasure from it''} (I9). In monetising their apps, developers found it important to take children's developmental needs into consideration:

\begin{quote}
    \textit{``We want to make sure that we also have a sensitive side where we take the users into account, and we don’t want them to be like, `oh I’ve gotta check my phone again, who knows if I have another like'.''} -- I18
\end{quote}



In attempts to monetise apps in an age-appropriate way, they faced several major challenges. First, apps not based on ads, i.e., paid apps, did not earn enough for developers to sustain their business or earn a sufficient income (15/20):

\begin{quote}
    \textit{``Well the first app, I made a free version, which only has about half of the elements to choose from. And then I made a paying version. Turns out, that [I] do not earn any money at all. I earned like 40 euros last year.''} -- I17
\end{quote}


Second, several interviewees and survey respondents discussed that users often revolted against paid apps, using rating systems to pressure developers to make them available for free (3/20):

\begin{quote}
    \textit{``You know, 100,000 people in Russia give you 1 star, just because they are used to seeing ads instead of paying money and you don’t offer this opportunity to them. This impacts your global rating; and this impacts your global sales in territories where you don’t show ads and are dependent on sales. [...] In order for us to make revenue, your game design has to be focused around showing ads.'''} -- I8 
\end{quote}

\begin{quote}
    \textit{``The primary reason to adopt child-safe advertising was an outrage on the app store comment/ratings that resulted in low ratings of our paid (premium) games just because it's paid. We were forced to react and offer some alternatives to appease those users, including the display of child-safe ads.''} -- S54
\end{quote}


Lastly, one discussed the cutthroat and sinister nature of marketplaces: even if you tried to use age-appropriate methods and became successful, it was likely that some clone-maker would duplicate your app and use privacy-invasive (but higher-yield) monetisation methods in it instead. That was particularly the case for paid apps, which were often cloned, turned into ad-based monetisation vehicles, and released for free (3/20):

\begin{quote}
\textit{I can find and show you dozens of apps that are clones of our games, so someone somewhere took our app, re-skinned it a little bit, and then left it as it is, with all the game design and whatever, and offered it on Google Play, and show all kinds of ads: safe, unsafe, whatever. […] So if this continues, essentially it means that users who are looking for free stuff and so on, they will still be able to find whatever they want. […]  It’s not even David vs Goliath, it’s a Flea vs Goliath. ''} -- I8
\end{quote}


As a consequence, many accepted that, while using ads did not align with their goal to limit data collection, they conceded there were few options, as these were a \textit{``necessary evil to keep the business afloat''} (I3) (14/20).
As a result, interviewees felt that commercial success of the app often came at a cost of the best interests of children (10/20), because \textit{``the two things don’t align very well, I suppose, and its kind of hard to do them both ways: to have an ethical thing and have the biggest payout in your app''} (I18). 

Several others expressed that whilst they tried methods other than advertising, sooner or later they had to resort to it in the end:
\begin{quote}
    \textit{``We were completely driven by the good for the end user. And then we ran out of money each time. And then we realised there has to be a balance of some degree of commercial success that will allow the app to sustain itself, and earn money to continue to operating and be as responsible for the end user.''} -- I10
\end{quote}

\begin{quote}
    \textit{``
    We didn’t want ads. If the market would allow us, if there was a possibility to still exist and do what you love, by offering paid apps, for 2.99, we would gladly remain in that field. Its just that the business there became impossible and unsustainable.''} -- I8
\end{quote}










\begin{table*}[]
\caption{Barriers and practices related to each of the three key objectives of our app developers. There where survey data is available, it is presented in terms of percentages and suffixed with the letter `S'.}
\small
\centering
\begin{tabular}{p{0.30\linewidth}p{0.30\linewidth}p{.30\linewidth}}
\hline
\multicolumn{1}{c}{\textbf{Goals}} & \multicolumn{1}{c}{\textbf{Constraints and Enablers}} & \multicolumn{1}{c}{\textbf{Practices}}  \\ \hline
\textbf{Design for children:} 
\begin{itemize}[noitemsep,leftmargin=*]
    \item Design for children is `different' than design for adults (15/20).
    \item Children's developmental needs are important to consider (10/20).
    \item Important to enforce age appropriate design features (11/20).
\end{itemize}
&
\textbf{Challenges with existing design guidelines:}
\begin{itemize}[noitemsep,leftmargin=*]
    \item Struggle to find children-specific design guidelines (11/20).
    \item Challenging to keep up with the constant change of guidelines (7/20).
\end{itemize}
&
\textbf{Use commercial guidelines and intuition:}
\begin{itemize}[noitemsep,leftmargin=*]
    \item Design based on following intuition (17/20; 63\%S).
    \item Try to focus on good user experience instead of addictive features (17/20).
    \item Follow Google's app publication guidelines (10/20).
    \item Consult professions to help with app design (3/20; 3\%S).
\end{itemize}
\\ \hline 

\textbf{Create good apps:}
\begin{itemize}[noitemsep,leftmargin=*]
    \item Data is needed to understand user behaviour and improve app (9/20; 6\%).
    \item Data collection is needed to main efficient core service (5/20).
    \item Need for third-party libraries to develop apps (13/20; 63\%S).
\end{itemize}
&
\textbf{Perception of children's data:}
\begin{itemize}[noitemsep,leftmargin=*]
    \item Collecting any data from children is wrong (8/20).
    \item Data collection is necessary and not always harmful (6/20).
    \item Children's data not useful (2/20).
    \item Collecting data poses legal risks (4/20).
\end{itemize}

\textbf{Opaque ecosystem:}
\begin{itemize}[noitemsep,leftmargin=*]

    \item Libraries and SDKs are largely opaque and lack of transparency (13/20).
    \item Limited capability to influence large cooperates' practices (6/20).
\end{itemize}
&
\textbf{Data collection through data controllers:}
\begin{itemize}[noitemsep,leftmargin=*]
    \item Avoid data collection if possible (6/20; 68\%S).
    \item Limit to anonymous data collection only (9/20).
    \item Trust libraries from prominent providers (11/20; 64\%S).
    \item Technical limitation to data collection (7/20).
    \item No data sharing with third parties (95\%S).
\end{itemize}
\\ \hline 
\textbf{Earn an income to sustain the business:}
\begin{itemize}[noitemsep,leftmargin=*]
    \item Need to earn enough income to continue development(13/20; 63\%S).
    \item Monetise in an age-appropriate way (17/20; 85\%S).
\end{itemize}
&
\textbf{Developer perceptions:}
\begin{itemize}[noitemsep]
    \item Apps themselves do not make enough money (15/20).
    \item Pressure from the end users (3/20).
    \item Free clones due to competition (3/20).
    \item Market success has to rely on ads (14/20).
    \item Commercial success comes at a cost (10/20; 21\%S).
\end{itemize}

&
\textbf{Use child-safe advertising based monetisation: }
\begin{itemize}[noitemsep,leftmargin=*]
    \item Advertisement based methods support app growth and access whilst sustaining adequate revenue streams (14/20).
    \item Lack of sustainable alternatives make this the only viable option in practice.
\end{itemize}
\\ \hline 

\end{tabular}

\label{tab:themes}
\normalsize
\end{table*}

\section{Discussion}
From our findings we identified three key challenges in designing privacy-friendly and age-appropriate apps for children: (i) navigating the complex and opaque landscape of third-party services and their underlying data economy, (ii) the lack of viable monetisation options not relying consumer data, and (iii) the lack of awareness and applicability of regulatory guidelines. An overview of this is also presented in Table \ref{tab:themes}. Below we discuss why these challenges interfere with best practices set out in data protection frameworks and how they fail to address these issues.

\subsection{The Challenges of Using Third-Party Libraries}




Our findings show that developers extensively use and rely on third-party libraries to provide core services and gain user insight. More specifically, our participants often relied on libraries from major data controllers, like Google Analytics (58.5\%), Google Firebase (22\%), and Facebook APIs (29.3\%). These findings are consistent with former research. Large scale analyses of Android apps has shown up to 82\% of these apps and 60\% of all code in these apps is from third-party libraries \cite{wang2015wukong, linares2014revisiting}. Use of third-party libraries has been shown to be critical in saving development time and efforts \cite{baldassarre2005industrial, lim1994effects, morisio2002quality}. 

While developers demonstrated to care about children's privacy, it also well known that third-party libraries can be a major gateway to global data tracking and profiling networks \cite{binns2018third, razaghpanah2018apps}. Products like Google Analytics and Firebase, which were popular amongst our participants, are free so that it can ``provide confidence and prove the value of online advertising to potential new advertisers" \cite{clifton_analytics}, meaning they rely on the profiling of large volumes of user data. These issues are also addressed by the data protection frameworks we analysed in section 3. The use of analytics and storage libraries relates to requirements about \textit{data sharing} and \textit{profiling} (see Table \ref{table_ico_codes}). Both the AADC and COPPA require that data should not be shared with third parties unless there is a compelling reason to do so and the AADC only allows profiling under the assurance that reasonable security measures are in place. 

This is not to say that app developers should never make use of such third-party libraries, but it does have important implications for efforts addressing `age-appropriate' design. Firstly, our findings show that developers have conflicting perceptions of major data controllers. They realise that, on the one hand, data controllers are nontransparent and potentially harmful in their data handling practices (13/20), whilst simultaneously relying on them to produce high-quality and unmalicious software (13/20). This forced trade-off highlights the complexity and lack of guidance in navigating the often treacherous landscape of third-party services. Data protection guidelines, such as the AADC and COPPA, do not provide sufficient support to developers in this regard. For example, they do not concretely state which libraries are simultaneously reliable and `age-appropriate'. 

Second, developers generally perceived the use of libraries such as Google Analytics and Firebase as harmless and non-detrimental to children's privacy (8/20). This demonstrates the lack of a clear definition of what `privacy-friendly', `age-appropriate', or `safe' third-services are supposed to be. While these boundaries are not clearly defined in data protection frameworks, both developers and regulators cannot make sound judgements in selecting or auditing third-party libraries. 

Lastly, developers also indicated that third-party libraries often behave in unpredictable and unknown ways (13/20). This demonstrates that even if developers have the best of intentions, they cannot always enforce their well intended principles of designing for children. The AADC and COPPA require developers to ensure that third-party services have reasonable protections in place, for example by examining their privacy policies. However, what they fail to take into account is the widely accepted fact that privacy policies are rarely read \cite{obar2020biggest} and that developers have difficulties understanding the legal language presented in them \cite{assal2019think, balebako2014privacy}. 

\subsection{Monetisation in a Data-Driven Economy}
Our participants expressed their aim to monetise their apps in a privacy friendly way without using advertisements. This sentiment is consistent with the key message in the data protection frameworks we examined. The AADC mentions personalised marketing and advertising in their \textit{profiling} section, stating that parental consent is required for behavioural advertising, as ``‘legitimate interests’ is unlikely to provide a valid lawful basis for processing for this purpose" \cite{ico_2020}. Similarly, Google and Apple do not allow third-party personalised advertising, with or without parental consent. Apple actually provides additional information for developers on business models and explains how best to implement these models \cite{apple_business_models}. However, they are geared towards maximising revenue by emphasising the importance of ``optimizing with analytics" and ``extensive user acquisition marketing campaigns" \cite{apple_freemium}, which is slightly misaligned the principles advocated by the AADC, COPPA, and COPFE.

A few developers tried to adhere to the principles set out above, by making use of non-advertisement based monetisation. However, they faced several barriers:

First, earning a sustainable income through subscriptions or pay-to-download business models, requires a large number of downloads, which is difficult to achieve for the average developer in a marketplace with over 3 million apps. Approximately 40\% of the apps never gain more than a handful of users \cite{sigg2019exploiting} and adding paywalls or making the app pay-to-use increases the threshold for a user to download it. It is also for this reason that a large number of our interviewees indicated that apps do not earn enough (10/20).

Second, merely having in-app purchases or subscription options is not enough. Participants expressed that earning money through in-app purchases or subscriptions also requires persuading users into making these purchases. At the same time, they felt morally conflicted in doing this.





Third, business models based on premium apps are not always accepted by the larger end user community. Several of our participants faced pressure to make their paid apps free, as their users could not afford to pay for it. Users either leave negative reviews or simply switch to a different app which can also fulfil their requirements. The competitive nature of the Google Playstore has put developers in a position to oblige with user's requests, and users generally want free apps. Developers are then forced to implement ads as the only alternative.

As a result, many developers in the interview study reported that they had to rely on in-app ads or sharing of data with third-parties to retain a business viability (14/20). This was further supported by our survey data, which shows that more than 79\% monetised their app. 

This phenomenon demonstrates the problem that guidelines do not address developers' need to monetise apps, nor do they provide enough guidance on this matter. First, while the need for advertising is evident, they do not specify which and when advertising networks are appropriate to use. This lack of regulation in the mobile advertising industry makes it challenging for developers to judge the boundaries of `safe' and `privacy-friendly' monetisation. Second, there is conflicting information between commercial guidelines, which focus on `user acquisition' and data processing, and principles set out by the AADC and COPFE, which is not discussed by any of the frameworks.

\subsection{Problems in Using Guidelines}
Finally, we identified the lack of awareness and accessibility to guidelines as a key challenge for developers (11/20). As a result, our participants reported to default to making use of terms and conditions set out by Google as a framework for child appropriate development (10/20). 

However, there are a few problems with relying on recommendations made by Google. First, Google's ``Designing Apps for Children and Families'' guidelines are not directly aligned with the AADC. Importantly, they do not adequately address \textit{data minimisation} principles which are core to the AADC. For example, the ``Ads and Monetization'' policy for children, by Google, is primarily focused on content based restrictions, such as `inappropriate ad content', `multiple ad placements', and `use of shocking or emotionally manipulative tactics' \cite{google_ads_policy}, instead of addressing any privacy implications of ad-based monetisation. 

Second, Google's guidelines and tutorials place their own products in the limelight, such as Google AdMob and Google Ad Manager \cite{google_approved_sdks}, thereby increasing the adoption rate of third-party libraries in children's apps by major data controllers. 

\subsection{Working Towards Solutions}

Our study has given us a detailed understanding of challenges developers face in designing privacy friendly and age-appropriate apps for children. Based on our findings and analyses, in this section we make concrete recommendations to address some of the problems highlighted above. 

\subsubsection{Include developers as stakeholders}
Our work highlights the importance of including app developers as stakeholders in creating design guidelines \cite{assal2019think}. It is widely recognised that design guidelines should reflect stakeholders' values~\cite{wisniewski2017parental,nouwen2015value} and be co-developed with key stakeholders~\cite{kumar2018co,hiniker2017co}, such as children and families. 

\subsubsection{The need for supporting documents}
In order to help developers navigate the complex app development ecosystem, organisations like the ICO need to provide clarity and specific requirements for what constitutes `age-appropriate' or `privacy-friendly libraries'. 

To expose developers to appropriate libraries other than those from Google, guidelines need concrete recommendations in select such services, for example through a knowledge base of resources and services approved by regulatory bodies.

\subsubsection{Additional tools for developers}
Many developers mentioned that they do not fully understand how data is handled by data controllers and market leaders, or the libraries they use behave in unpredictable ways. This highlights the need for additional support through practical tools and resources. There is an increasing amount of research supporting usable security tools for developers \cite{assal2019think,smith2015questions,xie2011aside}, however what is needed are resources to help developers navigate the vast choices of third-party libraries.

    \subsubsection{Need for industry support}
History has shown us that it takes a market leader to move the market. For example, in 2019 Apple restricted all third-party advertising and analytics. Similarly, their new iOS 14 operating system will require apps to explicitly ask its users permission to allow tracking services \cite{apple_guide}. This approach to technology design has impacted companies whose business model centres around data collection and sharing, such as Facebook \cite{apple_iphone_facebook}. For data tracking to come to an end on the Google Playstore, Google will have to push for similar policies. While the need for industry support is evident, it is unrealistic for Google to adopt such principles. Apple can afford to do this, because their advertising revenue is believed to only make up a small part of their total earnings \cite{newton_2020}. Google, on the other hand, produced 83.3\% of the 2019 revenue through the advertising industry \cite{google_advertising}, which is strongly grounded in the data tracking and profiling. 

While industry support is lacking, the HCI community plays in an important role in facilitating change in the ecosystem. For example, by raising awareness of privacy risks, making these risks more comprehensible, and facilitating the adoption of privacy preserving tools amongst (young) end users.

\subsubsection{Alternate monetisation methods}
The need for privacy friendly monetisation methods is already recognised by the wider community. For example, Apple has an initiative which facilitates privacy-friendly monetisation, called Apple Arcade \cite{apple_arcade}. End users have access to a set of apps through a monthly subscription, without in-app purchases or advertisements. While this is a good initiative, it is not a solution to the problem of data-driven monetisation. Apple Arcade only has about 100 games and studios were paid a fee for having their game included, which does not scale well to the large number of apps targeted at children. 

Similarly, business models which do not make use of ads, such as the `pay-to-download' business model, are not automatically privacy friendly. Paid services come with their own set of problems and ethical dilemmas. Firstly, it leads to the problem that privacy becomes a rather expensive commodity, creating digital exclusion and a lack of access to those who cannot afford it. It is widely accepted that marginalised groups \cite{vickery2015don} have diminished privacy, and commoditising privacy in a densely populated marketplace would only exacerbate the problem. Secondly, numerous studies have shown that paying for apps does not significantly reduce the amount of data tracking and collection \cite{bamberger2020can, han2019you, seneviratne2015measurement}. Paid apps still contain third-party tracking libraries and require dangerous permissions. The fact that there is no trivial solution to this, highlights the need for additional research in this area. 

    \subsubsection{Support for end users}
Lastly, it is also important to consider support for end-users. Research around end users has been focused on raising privacy awareness. However, research in raising users’ awareness of age-appropriate design implications is limited, which can be a key barrier to privacy-friendly technology innovation. By not depending on higher level policy changes by developers and market leaders, end users can exercise a finer level of control over their privacy choices. For example, there are technologies currently in existence which inform users of trackers in apps and also give them the ability to control how these apps behave. 
    
\subsection{Limitations}
We are aware of several limitations and methodological decisions which constrain our findings. Below we provide a rationale for this and suggest opportunities for future research which can build on our findings. 

\subsubsection{Recruitment}
In our recruitment we limited ourselves to the Android marketplace. The primary reason for this is that Android is the dominant operating system with a global market share of 75\%\footnote{https://gs.statcounter.com/os-market-share/mobile/worldwide}. Also, our research is motivated by the knowledge that Android applications are known to distribute data to third-parties on a large scale. While the presence of trackers and third-party libraries in Android applications is well researched \cite{binns2018third, wang2015wukong}, the privacy landscape of iOS apps is less documented. In our future work we aim to extend this study to include iOS apps and developers as well. 

Additionally, we also limited ourselves to the western app development market, focusing on US and European regulations. This choice is motivated by regulatory interventions taking place in Europe, primarily ICO's enforcement of the statutory AADC, which will come into effect in 2021. These changes will have an imminent impact on developers who wish to publish apps in the UK Google Playstore. 

\subsubsection{Study participants}

Since participation in our study was purely voluntary, the participants who responded may have represented those with a better awareness of potential issues or took children's best interests more seriously than the average developer. Although this may have skewed our findings, our focus on the gaps in current practices and knowledge still provided valuable insights into a research direction of which little is known.  

Similarly, our study captured only a small fragment of all Android developers. While the principles developers aimed to realise aligned with best practices, we do not claim that this is the case for all developers. It is likely that have malicious intents or are indifferent about children's privacy. However, we wanted to show that those developers who prioritise children's beneficence, well-being, and safety, are faced with real challenges and conflicts in supporting these priorities.

Lastly, we have not yet verified the features claimed to have been implemented by the participants by conducting a technical analysis of their apps (for example by analysing tracker domains the app contacts). This would not only help us to follow up conversations with our study participants to gain a deeper understanding of their awareness and knowledge, but also to validate the claims given by the participants.

\section{Conclusion}

Through 20 in-depth interviews and 134 surveys with children's app developers, our study provided a much-needed developers’ perspective regarding challenges and practices related to developing Android apps for children. We analysed the results of our data through the lens of 5 leading data protection frameworks. Our findings show that most developers were morally and legally motivated to try supporting children’s best interests. They indicated to make use of techniques to avoid engaging in unnecessary data collection and data sharing practices. 
However, at the same time, we also identified key barriers and practices which stood in the way of designing for children's best interests. Firstly, developers relied on third-party libraries from major data controllers for analytics and functionality. They often were not aware of alternatives or assumed them to be harmless to end users. Secondly, developers struggled to monetise their apps in privacy-friendly ways without relying on advertising networks. Market pressure and a lack of alternatives often forced them to adopt advertising as a primary source of revenue, thus making a trade-off between commercial success and children's privacy needs. Lastly, we identified a lack of awareness and accessibility to concrete and neutral design guidelines. Developers often relied on Google's app publication requirements, which are not always directly aligned with principles set out by the AADC. We call for (1) supporting documents to aid developers to navigate the opaque and complex development ecosystem, (2) research into alternative and privacy-friendly monetisation methods, (3) tools to support developers in auditing their apps, and (4) tools to empower end users in case developers do not keep up with age-appropriate practices. 

\bibliographystyle{ACM-Reference-Format}
\bibliography{bibliography}

\appendix

\clearpage

\begin{table*}[t]
\section{Interview participant demographics}
\vspace{1em}

\caption{\label{table:demographics}Table containing descriptive characteristics of the apps developed by our developers. App 4 was removed from the Google Play Store and App 18 was in development at the time of the interview. }
\begin{tabular}{lp{0.12\linewidth}lll}
\hline
\textbf{App} & \textbf{Country} & \textbf{Playstore Category} & \textbf{Age range} & \textbf{App description}                          \\ \hline
1            & Turkey           & Tool                        & PEGI 3             & Parental control.                                 \\ \hline
2            & United States    & Entertainment               & PEGI 3             & Parental control.                                 \\\hline
3            & Canada           & Parenting                   & PEGI 3             & Parental control.                                 \\\hline
4            & Bosnia           & -                           & -                  & Parental control.                                 \\\hline
5            & Canada           & Tools                       & PEGI 3             & Parental control.                                 \\\hline
6            & Austria          & Education                   & PEGI 3             & App teaches children programming.                 \\\hline
7            & Germany          & Education                   & Ages 6 - 12        & Teaches children mathematics.                     \\\hline
8            & Lithuania        & Educational, Pretend Play   & PEGI 3             & Teaches children about every day aspects of life. \\\hline
9            & Macedonia        & Casual, Brain Games         & PEGI 3             & Game based on reflexes.                           \\\hline
10           & Russia           & Education                   & Ages 8 \& under    & Educational games.                                \\\hline
11           & Russia           & Education                   & Ages 8 \& under    & Educational games.                                \\\hline
12           & Russia           & Education                   & Ages 8 \& under    & Educational games.                                \\\hline
13           & Bangladesh       & Education                   & PEGI 3             & Various tools for school.                         \\\hline
14           & Tunisia          & Education                   & PEGI 3             & Educational game to learn about health.           \\\hline
15           & United Kingdom   & Parenting, Education        & PEGI 3             & Augmented reality to enhance outdoors learning.   \\\hline
16           & Portugal         & Education                   & PEGI 3             & Assists in school planning.                       \\\hline
17           & Austria          & Education                   & PEGI 3             & Artistic game to compose drawings.                \\\hline
18           & The Netherlands  & -                           & -                  & -                                                 \\\hline
19           & Germany          & Education                   & Ages 6 - 12        & Media streaming for children.                     \\\hline
20           & Columbia         & Action                      & PEGI 3             & Action game.                                      \\ \hline
\end{tabular}

\end{table*}



\end{document}